\documentclass[10pt]{IEEEtran}

\usepackage{times}

\usepackage[dvipsnames]{xcolor}
\usepackage[normalem]{ulem}
\usepackage{lineno,hyperref}

\usepackage{graphicx}
\usepackage{subcaption}
\usepackage{bbold} 
\usepackage{amsmath,amssymb,mathrsfs,amsthm} 

\usepackage{algorithm}
\usepackage{algpseudocode}
\usepackage{multirow}
\usepackage{gensymb}
\usepackage{enumitem} 

\newtheorem{lemma}{Lemma}

\DeclareMathOperator*{\argmin}{arg\,min}

\usepackage{nomencl}
\makenomenclature

\begin{document}

\title{Maximum Marginal Likelihood Estimation of Phase Connections in Power Distribution Systems}
\date{}
\author{\IEEEauthorblockN{Wenyu Wang, Student Member, \textit{IEEE}, Nanpeng Yu, Senior Member, \textit{IEEE}}\\
}

\maketitle

\begin{abstract}
Accurate phase connectivity information is essential for advanced monitoring and control applications in power distribution systems. The existing data-driven approaches for phase identification lack precise physical interpretation and theoretical performance guarantee. Their performance generally deteriorates as the complexity of the network, the number of phase connections, and the level of load balance increase. In this paper, by linearizing the three-phase power flow manifold, we develop a physical model, which links the phase connections to the smart meter measurements. The phase identification problem is first formulated as a maximum likelihood estimation problem and then reformulated as a maximum marginal likelihood estimation problem. We prove that the correct phase connection achieves the highest log likelihood values for both problems. An efficient solution method is proposed by decomposing the original problem into subproblems with a binary least-squares formulation. The numerical tests on a comprehensive set of distribution circuits show that our proposed method yields very high accuracy on both radial and meshed distribution circuits with a combination of single-phase, two-phase, and three-phase loads. The proposed algorithm is robust with respect to inaccurate feeder models and incomplete measurements. It also outperforms the existing methods on complex circuits.
\end{abstract}

\begin{IEEEkeywords}
Distribution network, maximum marginal likelihood estimation, phase identification.
\end{IEEEkeywords}

\nomenclature[1I]{$I_{n}$}{$n\times n$ identity matrix.}
\nomenclature[1Im]{$Im(\cdot)$}{Imaginary part of a complex variable.}
\nomenclature[1Re]{$Re(\cdot)$}{Real part of a complex variable.}
\nomenclature[1M]{$M$}{Number of loads in a circuit.}
\nomenclature[1N]{$N$}{Number of three-phase non-substation nodes in the distribution network.}
\nomenclature[1Y]{$Y^{ij}$}{Bus admittance matrix between phase $i$ and $j$.}
\nomenclature[1diag]{$\textmd{diag}(\cdot)$}{$\textmd{diag}(\boldsymbol{x})$ of a vector $\boldsymbol{x}$ is a diagonal matrix with $\boldsymbol{x}$ on the main diagonal. $\textmd{diag}(X_1,...,X_n)$ is a block diagonal matrix with diagonal matrices of $X_1,...,X_n$.}
\nomenclature[1v]{$\boldsymbol{v}$, $\boldsymbol{\theta}$, $\boldsymbol{p}$, $\boldsymbol{q}$}{Vector of voltage magnitudes, voltage angles, real power injections, and reactive power injections of 3 phases of the nodes.}
\nomenclature[1v1]{$\boldsymbol{\check{v}}$, $\boldsymbol{\check{\theta}}$, $\boldsymbol{\check{p}}$, $\boldsymbol{\check{q}}$}{Non-substation nodes' voltage magnitude and angle difference with the substation, and their real and reactive power.}
\nomenclature[1v2]{$\boldsymbol{\hat{v}}$, $\boldsymbol{\hat{p}}$, $\boldsymbol{\hat{q}}$}{Vectors of load measurements of voltage magnitudes, real power, and reactive power injections.}
\nomenclature[1v3]{$\boldsymbol{\tilde{v}}$, $\boldsymbol{\tilde{p}}$, $\boldsymbol{\tilde{q}}$}{Time differenced load measurements of voltage magnitudes, real, and reactive power injections.}
\nomenclature[1v4]{$\boldsymbol{\overline{v}}$, $\boldsymbol{\overline{\theta}}$}{Flat voltage solution of three-phase power flow.}
\nomenclature[1v5]{$\boldsymbol{\hat{v}^{\text{ref}}}$}{Vector of reference voltage for loads.}
\nomenclature[1x1]{$x_m^i$}{Decision variable of load $m$'s phase connection.}
\nomenclature[1x2]{$\boldsymbol{x}$}{Vector of decision variables $x_m^i$.}
\nomenclature[1x3]{$\boldsymbol{x^*}$}{True value of the decision variable vector $\boldsymbol{x}$.}
\nomenclature[2alpha]{$\alpha$}{Rotation operator, $\alpha=e^{-j\frac{2\pi}{3}}$.}
\nomenclature[31]{$\mathbf{1}_n$}{An all-$1$ vector of size $n$.}
\nomenclature[3i]{$(\cdot)^i$}{A variable in phase $i$.}
\nomenclature[3i]{$(\cdot)^{ij}$}{A variable between phase $i$ and $j$.}
\nomenclature[3n]{$(\cdot)_n$}{A variable at node or load $n$.}
\nomenclature[3n]{$(\cdot)_{-n}$}{A variable excluding node or load $n$.}
\nomenclature[3t]{$\cdot (t)$}{The value of a variable at time $t$. }

\printnomenclature

\section{Introduction} \label{Sec_intro}
With declining costs, distributed energy resources (DERs) such as energy storage systems, distributed generation, and electric vehicles are rapidly penetrating power distribution systems around the world. To coordinate the operations of a large number of heterogeneous DERs, advanced distribution system control applications such as Volt-VAR control, network reconfiguration, and three-phase optimal power flow need to be implemented. The successful implementation of these applications requires accurate information about the phase connectivity of power distribution systems. However, the phase connectivity information in electric utilities is usually missing or highly unreliable. 

Traditionally, electric utilities send field crews to measure phase angles and determine phase connections with special equipment such as phase meters \cite{bierer2010long}. Although such practices provide very accurate phase connections information, they are very labor-intensive, time-consuming, and expensive. The time synchronized measurements from micro-phasor measurement units ($\mu$PMUs) can also provide highly accurate estimations of phase connections  \cite{wen2015phase,liao2018unbalanced}. However, a system-wide installation is cost prohibitive. State estimation can also be used to verify phase connection information \cite{krsman2017verification}. However, this method only applies to circuits with mostly accurate phase connections and the area of incorrect phase connections needs to be known. In order to develop more cost effective phase identification algorithms, researchers have turned to data-driven methods, which use measurements from the advanced metering infrastructure (AMI). The existing data-driven approaches can be categorized into three approaches: energy supply and consumption matching, correlation-based analysis, and clustering-based analysis.

The energy supply and consumption matching approach is based on the principle of conservation of energy. With complete coverage of load measurements, the aggregate power consumption of downstream loads in each phase plus losses is equal to the corresponding phase's power flow measured at the upstream point. In this approach, Ref. \cite{dilek2000integrated} formulates the problem as integer programming and solves it using tabu search. Ref. \cite{kumar2017leveraging} uses relaxed integer programming and improves the phase identification accuracy by actively managing the power injections of DERs. In \cite{pappu2018identifying}, principal component analysis (PCA) and its graph-theoretic interpretation are used to infer phase connections. However, algorithms in this approach cannot identify phase connections in the presence of delta-connected two-phase loads.


In the correlation-based analysis approach, correlation analysis is performed using smart meters' and the substation's measurements or the three-phase primary line's measurements. Each smart meter is assigned to a phase, which has the highest correlation coefficient with it. In this approach, Ref. \cite{short2013advanced,luan2015smart} use voltage magnitude profiles for the correlation analysis. In \cite{xu2018phase}, salient features are extracted from load profiles for the correlation analysis. Although the correlation-based analysis has achieved good performance on radial circuits with only single-phase loads, it does not work well for a meshed circuit, which has all seven possible phase connections of single-phase, two-phase, and three-phase loads.


In the clustering-based approach, smart meters are grouped based on the mutual similarity of their voltage magnitude profiles. It is assumed that each resulting cluster represents a single phase connection. Ref. \cite{wang2016phase,wang2017advanced} project the voltage magnitude profiles onto low-dimension spaces and leverage constrained clustering algorithms to identify both single-phase and two-phase connections. Ref. \cite{olivier2018phase} designs an algorithm by combining clustering and the minimum spanning tree method to identify phase connections. However, it has been shown that the performance of the clustering-based approach deteriorates as the feeder becomes more balanced \cite{wang2017advanced}.

To further improve the phase identification accuracy and provide a theoretical foundation for the problem, we develop a physically inspired machine learning method for phase identification. By linearizing the three-phase power flow manifold, we first develop a physical model, which links phase connections to the smart meter measurements. We then formulate the phase identification task as a maximum likelihood estimation (MLE) problem and prove that the correct phase connection yields the highest log likelihood value. The nonlinearity and nonconvexity nature of the MLE problem makes it difficult to solve. Thus, we reformulate the MLE problem as a maximum marginal likelihood estimation (MMLE) problem and prove that the correct phase connection also yields the highest marginal log likelihood value. Finally, an efficient solution algorithm is developed for the MMLE problem by dividing it into sub-problems, which can be solved by least squares integer programming.


Compared to the existing data-driven phase identification algorithms, our approach has the following advantages: first, the physically interpretable MMLE formulation brings a solid theoretical foundation to the phase identification problem; second, our proposed algorithm not only works for radial distribution feeders, but also heavily meshed networks; third, our proposed algorithm achieves higher accuracy for complex circuits with both single-phase and two-phase connections and a lower level of unbalance, which create a lot of problems to existing data-driven methods; fourth, our proposed algorithm is robust with respect to inaccurate feeder models and incomplete measurements.


The rest of the paper is organized as follows. Section \ref{LinearModel} covers the problem setup and the linearized three-phase power flow model. Section \ref{Sec_model_for_phaseID} derives the model that links the phase connections to the smart meter measurements. Section \ref{Sec_MLE_form} formulates the phase identification problem as an MLE and MMLE problem and presents an efficient solution algorithm. A comprehensive numerical test is performed in Section \ref{Sec_numerical} to evaluate the performance of the proposed MMLE-based phase identification method. Section \ref{Sec_conclusion} states the conclusion.



\section{Problem Setup and Linearized Three-Phase Power Flow Model} \label{LinearModel}
\subsection{Problem Setup}
\begin{figure}[htb]
\centering
\includegraphics*[width=0.4\textwidth]{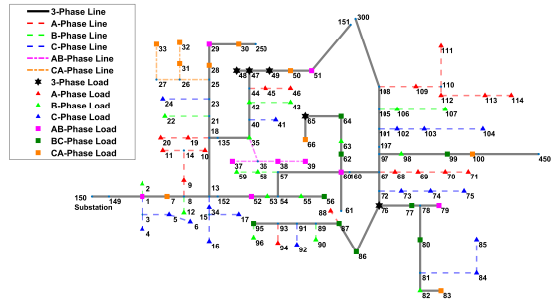}
\caption{Schematic of a modified IEEE 123-node test feeder.} \label{fig_123_node}
\end{figure}

We intend to identify the type of phase connection for all loads on a distribution feeder. The distribution feeder's three-phase primary line contains $N+1$ nodes, indexed as node $0$ to $N$, in which node $0$ is the source/substation. A load can connect to a three-phase node directly, or indirectly through a single-phase or two-phase branch (e.g., the dashed lines and dash-dot lines in Fig. \ref{fig_123_node}). Note that nodes and loads are two different concepts. In the technical derivation, all variables are in per unit or radian angles unless otherwise specified.

\subsection{Assumptions} \label{Assumptions}
Note that the assumptions described below are only used to prove that the correct phase connection yields the highest log likelihood value of the MLE and MMLE problem formulated in this paper. Some of these assumptions may not hold in the real world. However, the numerical study will show that our proposed algorithm still works well even when some of these assumptions no longer hold. In these cases, we can no longer guarantee that our proposed algorithm will result in $100\%$ accurate phase identification results.
\subsubsection{Data and Model Availability} First, the information about whether the load is single-phase, two-phase, or three-phase is assumed to be available. Usually, this information can be deduced by examining the distribution transformer configuration and customer billing information. Second, for a single-phase load on phase $i$, we know its power injection (both real and reactive power) and voltage magnitude of phase $i$. Third, for a two-phase delta-connected load between phase $i$ and $j$, we know its power injection and voltage magnitude across phase $i$ and $j$. Fourth, for a three-phase load, we know its total power injection and the voltage magnitude of one of the phases, which needs to be identified. Fifth, for the source node, we know the voltage measurement. Sixth, the connectivity model and the parameters of the primary feeder are known. Finally, we assume that the distribution feeder is not severely unbalanced. The task of phase identification is to determine which phase(s) each single-phase or two-phase load connects to and which phase's voltage magnitude the three-phase smart meter measures. Note that our proposed algorithm does not assume a 100\% smart meter penetration rate. The numerical study will show that our algorithm is robust with respect to incomplete measurements.



\subsubsection{Statistical Assumptions} First, it is assumed that the incremental changes in measured real, reactive power, and voltage magnitudes across one time interval are independent over time. Second, it is assumed that the noise terms which represent the model errors and the measurement errors are i.i.d. Gaussian. Note that the noise terms will be derived later in Section \ref{Sec_MLE_form}. Third, it is assumed that theses noise terms are independent of the incremental changes in smart meter measurements. Note that these statistical assumptions will be verified in the numerical study section.



\subsection{The Linearized Power Flow Model for Primary Feeders}

The very first step of our phase identification framework is to build a three-phase power flow model for the primary feeder. To do so, we need a procedure that we call \textit{reduction}, and the resulting network is called a \textit{reduced network}. The reduction is simply converting any loaded single-phase or two-phase branch into an equivalent load so that the reduced network contains only three-phase lines. The details of the reduction procedure is explained in Appendix \ref{1p2pEst}. In the rest of the paper, we use $M$ to denote the number of loads in the reduced network and \textit{load} refers to the equivalent load in the reduced network.

From the reduced primary feeder, by following \cite{bolognani2015fast}, we can derive the linearized three-phase power flow model shown in \eqref{eqLM-1}, with the variables organized by phase. The linearized model ignores shunt admittance because it is very small. Numerical study results will verify that ignoring shunt admittance does not affect the phase identification accuracy.

\begin{equation}\label{eqLM-1}
A
\begin{bmatrix}
       \boldsymbol{v-\overline{v}} \\
       \boldsymbol{\theta- \overline{\theta}}
     \end{bmatrix}
=
\begin{bmatrix}
A_{11} & A_{12} \\
A_{21} & A_{22}
\end{bmatrix}
\begin{bmatrix}
       \boldsymbol{v-\overline{v}} \\
       \boldsymbol{\theta-\overline{\theta}}
     \end{bmatrix}
=
\begin{bmatrix}
       \boldsymbol{p} \\
       \boldsymbol{q}
     \end{bmatrix}
\end{equation}
Here $A_{11}$, $A_{12}$, $A_{21}$, and $A_{22}$ are $3(N+1)\times 3(N+1)$ matrices. $\boldsymbol{v}$, $\boldsymbol{\theta}$, $\boldsymbol{p}$, and $\boldsymbol{q}$ are the nodes' voltage magnitude, voltage angle, and real and reactive power of three phases. $\boldsymbol{\overline{v}}=\mathbf{1}_{3(N+1)}$ and $\boldsymbol{\overline{\theta}}=[0\times \mathbf{1}^T_{N+1},-\frac{2\pi}{3}\times\mathbf{1}^T_{N+1}, \frac{2\pi}{3}\times\mathbf{1}^T_{N+1}]^T$ are the flat feasible solution for the underlying nonlinear power flow model. Let $\alpha=e^{-j\frac{2\pi}{3}}$, define $\Phi\triangleq \textmd{diag}(I_{(N+1)},\ \alpha I_{(N+1)},\ \alpha^2 I_{(N+1)})$ and define
\begin{equation}\label{eqLM-4}
Y \triangleq
\begin{bmatrix}
Y^{aa} & Y^{ab} & Y^{ac} \\
Y^{ba} & Y^{bb} & Y^{bc} \\
Y^{ca} & Y^{cb} & Y^{cc}
\end{bmatrix}
\end{equation}
where $Y^{ij}$ is the $(N+1) \times (N+1)$ nodal admittance matrix between phase $i$ and $j$. Then $A_{11}$, $A_{12}$, $A_{21}$, and $A_{22}$ can be calculated as $A_{11} =-A_{22}=Re(\Phi^{-1}Y\Phi)$ and $A_{12}=A_{21}=-Im(\Phi^{-1}Y\Phi)$.

It has been shown in \cite{kettner2019properties} that for a connected three-phase network, $\text{rank}(Y)=3N$. Thus, $\text{rank}(A)$ is at most $6N$. For subsequent derivations, we need to transform $A$ into a nonsingular form. Following Appendix \ref{NS_derive}, the transformed power flow model becomes
\begin{equation}\label{eqNS-16}
\check{A}
\begin{bmatrix}
\boldsymbol{\check{v}} \\
\boldsymbol{\check{\theta}}
\end{bmatrix}
=
\begin{bmatrix}
\check{A}_{11} & \check{A}_{12}\\
\check{A}_{21} & \check{A}_{22}
\end{bmatrix}
\begin{bmatrix}
\boldsymbol{\check{v}} \\
\boldsymbol{\check{\theta}}
\end{bmatrix}
=
\begin{bmatrix}
\boldsymbol{\check{p}} \\
\boldsymbol{\check{q}}
\end{bmatrix}
\end{equation}
where $\check{A}_{mn}$ is a $3N\times 3N$ matrix obtained by removing the rows and columns corresponding to the substation node in $A_{mn}$. We denote the difference of voltage magnitudes and voltage angles between the non-substation nodes and the substation nodes as $\boldsymbol{\check{v}}$, $\boldsymbol{\check{\theta}}$. We denote the non-substation nodes' real and reactive power as $\boldsymbol{\check{p}}$ and $\boldsymbol{\check{q}}$. 

In theory, $\check{A}$ is not guaranteed to be invertible. However, for the majority of real-world distribution feeders, $\text{rank}(\check{A})=6N$. It will be shown in the numerical study section that for all IEEE distribution test feeders, $\check{A}$ has a full rank.


Solving for ${\check{v}}$ with ${\check{p}}$ and ${\check{q}}$ from \eqref{eqNS-16}, we have
\begin{equation}\label{eqNS-22}
\begin{aligned}
\boldsymbol{\check{v}} = & (\check{A}_{11}-\check{A}_{12} \check{A}_{22}^{-1}\check{A}_{21})^{-1}\boldsymbol{\check{p}} \\
& - (\check{A}_{11}-\check{A}_{12} \check{A}_{22}^{-1}\check{A}_{21})^{-1}\check{A}_{12}\check{A}_{22}^{-1}\boldsymbol{\check{q}}
\end{aligned}
\end{equation}
or in condensed form as
\begin{equation}\label{eqNS-23}
\boldsymbol{\check{v}}=K\boldsymbol{\check{p}}-L\boldsymbol{\check{q}}
\end{equation}
It can be shown that $(\check{A}_{11}-\check{A}_{12} \check{A}_{22}^{-1}\check{A}_{21})$ is invertible if $\check{A}$ is invertible. Similarly, we can link $\boldsymbol{\check{\theta}}$ with $\boldsymbol{\check{p}}$ and $\boldsymbol{\check{q}}$ as
\begin{equation}\label{eqNS-24}
\begin{aligned}
\boldsymbol{\check{\theta}}= &(\check{A}_{12}-\check{A}_{11} \check{A}_{21}^{-1}\check{A}_{22})^{-1}\boldsymbol{\check{p}}\\
&-(\check{A}_{12}-\check{A}_{11} \check{A}_{21}^{-1}\check{A}_{22})^{-1}\check{A}_{11}\check{A}_{21}^{-1}\boldsymbol{\check{q}}
\end{aligned}
\end{equation}
or in condensed form as
\begin{equation}\label{eqNS-25}
\boldsymbol{\check{\theta}}=\mathcal{K}\boldsymbol{\check{p}}-\mathcal{L}\boldsymbol{\check{q}}
\end{equation}

\section{Model for Phase Identification} \label{Sec_model_for_phaseID}
In this section, we develop a mathematical model that relates the phase connections of loads to voltage magnitude and power injection measurements. Section \ref{CV2P} explains how to express smart meter measurements in terms of nodal voltages and power injections of the three-phase power flow model. Section \ref{Sec_integrate_x} derives the phase connection model, which relates phase connections to network measurements.


\subsection{Link Smart Meter Measurements with the Nodal Voltages and Power Injections}\label{CV2P}
The linearized three-phase power flow models \eqref{eqNS-23} and \eqref{eqNS-25} are derived in terms of nodal voltages and power injections $\boldsymbol{\check{v}}$, $\boldsymbol{\check{\theta}}$, $\boldsymbol{\check{p}}$, and $\boldsymbol{\check{q}}$, which are often not directly measured by smart meters. Thus, we need to embed the smart meter measurements into these two equations. This is straightforward for single-phase and three-phase loads. For a single-phase load $m$ on node $n$, its voltage measurement $\hat{v}_m$ is equal to one of the three phase-to-neutral voltage magnitudes $v_n^i \ (i=a,b,c)$, which is related to $\check{v}_n^i$ in \eqref{eqNS-16} via $\check{v}_n^i \triangleq v_n^i-v_0^i$, where $v_0^i$ is the source voltage magnitude in phase $i$. Similarly, a single-phase load's power injection measurement $\hat{p}_m+j\hat{q}_m$ corresponds to the power injection of one of the three phases $\check{p}_n^i+j\check{q}_n^i$ at node $n$. For a three-phase load $m$ at node $n$, the single-phase voltage measurement $\hat{v}_m$ is equal to one of the three nodal voltage magnitudes $v_n^i \ (i=a,b,c)$. We can assume that the three-phase power injections $\hat{p}_m+j\hat{q}_m$ is distributed relatively evenly to three phases at node $n$. For a delta-connected two-phase load, we need the following derivations to link its measurements to the three-phase power flow model.


\subsubsection{Link Power Injection Measurements with Power Flow Model}
Without loss of generality, we use a phase $AB$ load as an example. Suppose the two-phase power injection measurement is $S_{ab}=P_{ab}+jQ_{ab}=S_a+S_b=(P_a+jQ_a)+(P_b+jQ_b)$. Here, $S_a$ and $S_b$ are the power injections at the phase $A$ and phase $B$ ports. We can estimate $S_a$ and $S_b$ based on $S_{ab}$ as follows: (see the proof in Appendix \ref{App_Delta_power})
\begin{equation}\label{eq2P-1-1}
\begin{aligned}
S_a & \approx \bigg( \frac{1}{2}P_{ab} + \frac{\sqrt{3}}{6} Q_{ab} \bigg)+j\bigg( \frac{1}{2} Q_{ab} -  \frac{\sqrt{3}}{6} P_{ab} \bigg)
\end{aligned}
\end{equation}
\begin{equation}\label{eq2P-1-2}
S_b \approx \bigg( \frac{1}{2}P_{ab} - \frac{\sqrt{3}}{6} Q_{ab} \bigg)+j\bigg( \frac{1}{2} Q_{ab} +  \frac{\sqrt{3}}{6} P_{ab} \bigg)
\end{equation}

\subsubsection{Link Voltage Magnitude Measurements with Power Flow Model}
Here we need to establish a relationship between the phase-to-phase voltage magnitude measurements and the nodal phase-to-neural voltage magnitudes in \eqref{eqNS-23} and \eqref{eqNS-25}. For a load $m$ across phase $ij$ ($ij \in \{ab,bc,ca\}$) at node $n$, the relationship can be written as: (see the proof in Appendix \ref{App_2phase})
\begin{equation}\label{eq2P-10}
\begin{aligned}
\hat{v}_m-v^{ij}_0 \approx & \frac{\sqrt{3}}{2} (v^i_n-v^i_0) +\frac{\sqrt{3}}{2} (v^j_n-v^j_0) \\
& + \frac{1}{2} (\theta^i_n-\theta^i_0) -\frac{1}{2} (\theta^j_n-\theta^j_0)
\end{aligned}
\end{equation}
where $\hat{v}_m$ is load $m$'s voltage magnitude measurement. $v^{ij}_0$ is the voltage magnitude across phase $ij$ at the substation. $v_n^i$ and $v_0^i$ are the voltage magnitudes of phase $i$ at node $n$ and the substation. $\theta_n^i$ and $\theta_0^i$ are the voltage angles of phase $i$ at node $n$ and the substation. Note that in above derivations, voltages are in per unit and angles are in radian.

\subsection{Modeling Phase Connections in Three-phase Power Flow} \label{Sec_integrate_x}
\subsubsection{Decision Variables for Phase Connections} \label{Decision}
We use three decision variables, $x_m^1$, $x_m^2$, and $x_m^3$ to denote the phase connection for each load $m$. $x_m^i=0 \ \text{or} \ 1$, and $\sum_i x_m^i=1, \ \forall \ m$. If load $m$ is single-phase, then $x_m^1$, $x_m^2$, and $x_m^3$ represent $AN$, $BN$, and $CN$ connections. If $m$ is two-phase, then $x_m^1$, $x_m^2$, and $x_m^3$ represent $AB$, $BC$, and $CA$ connections. If $m$ is three-phase, and the measured voltage is between one phase and the neutral, then $x_m^1$, $x_m^2$, and $x_m^3$ represent which of the phases $AN$, $BN$, and $CN$ is measured. As stated in the assumptions, we know whether a load is single-phase, two-phase, or three-phase from the distribution transformer configuration and customer billing information. The phase connection decision variables form an $M\times 3M$ matrix $X$ defined as $X \triangleq \textmd{diag}([x_1^1 \ x_1^2 \ x_1^3],...,[x_M^1 \ x_M^2 \ x_M^3])$.

\subsubsection{Additional Definitions} \label{Def}
Several matrices and variables are defined here to build the model for phase connections.

Define matrices $W_1$ and $W_2$ as
\begin{equation}\label{eqML-2}
W_1 \triangleq
\begin{bmatrix}
1 & 1 & 0 \\
0 & 1 & 1 \\
1 & 0 & 1
\end{bmatrix}, \
W_2 \triangleq
\begin{bmatrix}
1 & -1 & 0 \\
0 & 1 & -1 \\
-1 & 0 & 1
\end{bmatrix}
\end{equation}

Let $I_n$ denote an identity matrix of size $n$, $\mathbb{0}_{k\times l}$ denote a $k \times l$ all-0 matrix, and $\mathbb{1}_{k\times l}$ denote a $k \times l$ all-1 matrix. Define $U^1$ and $U^2$ as $3M \times 3N$ matrices of $3\times 3$ blocks. Define $\hat{U}^1$ and $\hat{U}^2$ as $3N \times 3M$ matrices of $3\times 3$ blocks. Define $U^1_{mn}$ and $U^2_{mn}$ as the $mn$-th block of $U^1$ and $U^2$. Define $\hat{U}^1_{nm}$ and $\hat{U}^2_{nm}$ as the $nm$-th block of $\hat{U}^1$ and $\hat{U}^2$. If load $m$ is not connected to node $n$, then $U^1_{mn}$, $U^2_{mn}$, $\hat{U}^1_{nm}$, and $\hat{U}^2_{nm}$ are equal to $\mathbb{0}_{3\times 3}$. If load $m$ is connected to node $n$, then $U^1_{mn}$, $U^2_{mn}$, $\hat{U}^1_{nm}$, and $\hat{U}^2_{nm}$ are defined based on load $m$'s phase connection type, as shown in Table \ref{table_block_value}.
\begin{table}[htb]
\centering
\caption{Values of $3\times 3$ Blocks by Phase Connection Type if Load $m$ is Connected to Node $n$} 
\label{table_block_value}
\begin{tabular}{ c  c  c  c  c }
  \hline
  \hline
  Load $m$'s Phase Connection Type & $U^1_{mn}$ & $U^2_{mn}$ & $\hat{U}^1_{nm}$ & $\hat{U}^2_{nm}$ \\ \hline
  single-phase & $I_3$ & $\mathbb{0}_{3\times 3}$ & $I_3$ & $\mathbb{0}_{3\times 3}$\\ 
  two-phase & $\frac{\sqrt{3}}{2}W_1$ & $\frac{1}{2}W_2$ & $\frac{1}{2}W_1^T$ & $\frac{\sqrt{3}}{6}W_2^T$ \\
  three-phase & $I_3$ & $\mathbb{0}_{3\times 3}$ & $\frac{1}{3}\mathbb{1}_{3\times 3}$ & $\mathbb{0}_{3\times 3}$ \\ \hline \hline
\end{tabular}
\end{table}

Define $\boldsymbol{\hat{v}}^{\text{ref}} \triangleq [\boldsymbol{\hat{v}}^{\text{ref}}_1,\ldots,\boldsymbol{\hat{v}}^{\text{ref}}_M]^T$, where $\boldsymbol{\hat{v}}^{\text{ref}}_m=[v_0^a, v_0^b, v_0^c]$ if load $m$ is single-phase or three-phase; $\boldsymbol{\hat{v}}^{\text{ref}}_m=[v_0^{ab}, v_0^{bc}, v_0^{ca}]$ if load $m$ is two-phase. Here, $v_0^i$ denotes the substation's voltage magnitude of phase $i$, and $v_0^{ij}$ denotes the substation's voltage magnitude across phase $ij$.

\subsubsection{Phase Connection Model}
Now we can build the model, which links phase connections with the smart meter measurements. Let $\boldsymbol{\hat{v}}$, $\boldsymbol{\hat{p}}$, and $\boldsymbol{\hat{q}}$ be $M\times 1$ vectors of measured voltage magnitudes, real power, and reactive power of the $M$ loads. From \eqref{eq2P-1-1} - \eqref{eq2P-10}, Section \ref{Decision}, and \ref{Def}, we have:

\begin{equation}\label{eqML-4-7}
\boldsymbol{\check{p}} \approx \hat{U}^1 X^T \boldsymbol{\hat{p}} + \hat{U}^2 X^T \boldsymbol{\hat{q}}
\end{equation}
\begin{equation}\label{eqML-4-8}
\boldsymbol{\check{q}} \approx -\hat{U}^2 X^T \boldsymbol{\hat{p}} + \hat{U}^1 X^T \boldsymbol{\hat{q}}
\end{equation}
\begin{equation}\label{eqML-4-9}
\boldsymbol{\hat{v}} \approx X \boldsymbol{\hat{v}}^{\text{ref}} + X U^1 \boldsymbol{\check{v}} + X U^2 \boldsymbol{\check{\theta}}
\end{equation}
With a slight abuse of notations, the entries of $\boldsymbol{\check{p}}$, $\boldsymbol{\check{q}}$, $\boldsymbol{\check{v}}$, and $\boldsymbol{\check{\theta}}$ are organized by node in \eqref{eqML-4-7}-\eqref{eqML-4-9} (instead of by phase as in \eqref{eqNS-23} and \eqref{eqNS-25}). Equations \eqref{eqML-4-7} and \eqref{eqML-4-8} map the measured power injection of each load to the corresponding nodal power injections in the linearized power flow model. Take load $m$ connected to node $n$ as an example and suppose $x_m^1=1$. If load $m$ is single-phase, then its power injection is mapped to phase $A$ at node $n$. If load $m$ is two-phase, then its power injection is distributed to phase $A$ and $B$ at node $n$ according to \eqref{eq2P-1-1} and \eqref{eq2P-1-2}. If load $m$ is three-phase, then its power injection is evenly distributed to all three phases of node $n$.


Equation \eqref{eqML-4-9} links the voltage measurement $\boldsymbol{\hat{v}}$ with $\boldsymbol{\check{v}}$ and $\boldsymbol{\check{\theta}}$, i.e., the nodal line-to-neutral voltage magnitude and angle difference with the substation in the linearized power flow model. Take load $m$ connected to node $n$ as an example and suppose $x_m^1=1$. If load $m$ is single-phase or three-phase, then \eqref{eqML-4-9} can be reduced to $\hat{v}_m=v_0^a+(v_n^a-v_0^a)$, where $v_n^a$ is node $n$'s voltage magnitude in phase $A$. If load $m$ is two-phase, then \eqref{eqML-4-9} is equivalent to \eqref{eq2P-10}.




Substituting \eqref{eqNS-23}, \eqref{eqNS-25}, \eqref{eqML-4-7} and \eqref{eqML-4-8} into \eqref{eqML-4-9} yields
\begin{equation}\label{eqML-4-10}
\begin{aligned}
\boldsymbol{\hat{v}}
\approx & X \boldsymbol{\hat{v}}^{\text{ref}} + X\hat{K}X^T \boldsymbol{\hat{p}} + X\hat{L}X^T \boldsymbol{\hat{q}}
\end{aligned}
\end{equation}
where $\hat{K} \triangleq [(U^1 K+U^2 \mathcal{K})\hat{U}^1+(U^1 L+U^2 \mathcal{L})\hat{U}^2]$ and $\hat{L} \triangleq [(U^1 K+U^2 \mathcal{K})\hat{U}^2-(U^1 L+U^2 \mathcal{L})\hat{U}^1]$. Here, with a slight abuse of notations, $K$, $L$, $\mathcal{K}$, and $\mathcal{L}$'s entries are organized by node (instead of by phase as in \eqref{eqNS-23} and \eqref{eqNS-25}). Thus, \eqref{eqML-4-10} provides the physical model, which relates power injection measurements and phase connections to voltage magnitude measurements.


To remove trends and seasonality in time series data, we define the difference of the voltage measurement and its lagged variable as $\boldsymbol{\tilde{v}}(t)$, with $\boldsymbol{\tilde{v}}(t)\triangleq \boldsymbol{\hat{v}}(t)-\boldsymbol{\hat{v}}(t-1)$. $\boldsymbol{\tilde{v}}^{\text{ref}}(t)$, $\boldsymbol{\tilde{p}}(t)$, and $\boldsymbol{\tilde{q}}(t)$ are defined in a similar way. Thus, we have the time difference version of the physical model:
\begin{equation}\label{eqML-6}
\boldsymbol{\tilde{v}}(t)= X \boldsymbol{\tilde{v}}^{\text{ref}}(t) + X\hat{K}X^T \boldsymbol{\tilde{p}}(t) + X\hat{L}X^T \boldsymbol{\tilde{q}}(t) +\boldsymbol{n}(t)
\end{equation}
where $\boldsymbol{n}(t)$ is the ``noise term'' representing the error of the linearized power flow model, the measurement error, and all the other sources of noise not considered. In \eqref{eqML-6}, $\boldsymbol{\tilde{v}}(t)$, $\boldsymbol{\tilde{p}}(t)$, $\boldsymbol{\tilde{q}}(t)$, and $\boldsymbol{\tilde{v}}^{\text{ref}}(t)$ can be calculated from the smart meter and substation measurements. $\hat{K}$ and $\hat{L}$ can be derived from the feeder model. Thus, the task of phase identification is to estimate the phase decision variables in $X$.



\section{Maximum Marginal Likelihood Estimation of Phase Connections}\label{Sec_MLE_form}
In this section, we first formulate phase identification as an MLE problem and then as an MMLE problem. Next, we prove that the correct phase connection is a global optimizer of the MMLE problem. Lastly, we develop a computationally efficient algorithm to solve the MMLE problem.
\subsection{MLE Problem Formulation} \label{subsection_MLE}
Let $\boldsymbol{x}\triangleq[x_1^1,x_1^2,x_1^3,...,x_M^1, x_M^2, x_M^3]^T$ be the phase connection decision variable vector. Define $\boldsymbol{\tilde{v}}(t,\boldsymbol{x})$ as the theoretical differenced voltage measurement $\boldsymbol{\tilde{v}}(t)$ with phase connection $\boldsymbol{x}$:
\begin{equation}\label{eqPF-1}
\boldsymbol{\tilde{v}}(t,\boldsymbol{x}) \triangleq X \boldsymbol{\tilde{v}}^{\text{ref}}(t) + X\hat{K}X^T \boldsymbol{\tilde{p}}(t) + X\hat{L}X^T \boldsymbol{\tilde{q}}(t)
\end{equation}

Then $\boldsymbol{\tilde{v}}(t)=\boldsymbol{\tilde{v}}(t,\boldsymbol{x})+\boldsymbol{n}(t)$, where $\boldsymbol{x}$ is the phase connection decision variable vector that we need to estimate.

As stated in Section \ref{Assumptions}, we assume that the noise $\boldsymbol{n}(t)$ is independent of $\boldsymbol{\tilde{v}}^{\text{ref}}(t)$, $\boldsymbol{\tilde{p}}(t)$, and $\boldsymbol{\tilde{q}}(t)$ and is i.i.d. Gaussian $\boldsymbol{n}(t)\sim \mathcal{N}(\mathbb{0}_{M\times 1},\Sigma_n)$, where $\Sigma_n$ is an unknown underlying covariance matrix. Given these conditions, $\boldsymbol{n}(t)$ is also independent of $\boldsymbol{\tilde{v}}(t,\boldsymbol{x})$. Thus, the likelihood of observing $\{\boldsymbol{\tilde{v}}(t)\}^{T}_{t=1}$ given $\{\boldsymbol{\tilde{v}}^{\text{ref}}(t)\}^{T}_{t=1}$, $\{\boldsymbol{\tilde{p}}(t)\}^{T}_{t=1}$, and $\{\boldsymbol{\tilde{q}}(t)\}^{T}_{t=1}$ is a function of $\boldsymbol{x}$:
\begin{equation}\label{eqPF-2}
\begin{aligned}
& Prob(\{\boldsymbol{\tilde{v}}(t)\}^{T}_{t=1} | \{\boldsymbol{\tilde{v}}^{\text{ref}}(t)\}^{T}_{t=1}, \{\boldsymbol{\tilde{p}}(t)\}^{T}_{t=1}, \{\boldsymbol{\tilde{q}}(t)\}^{T}_{t=1};\boldsymbol{x}) = \\
& \frac{|\Sigma_n|^{-\frac{T}{2}}}{(2 \pi)^{\frac{MT}{2}}} \! \times \! \exp \! \Big\{\! -\!\frac{1}{2} \! \sum_{t=1}^T [\boldsymbol{\tilde{v}}(t)\!-\!\boldsymbol{\tilde{v}}(t,\boldsymbol{x})]^T \Sigma^{-1}_n [\boldsymbol{\tilde{v}}(t)\!-\!\boldsymbol{\tilde{v}}(t,\boldsymbol{x})]\!\Big\}
\end{aligned}
\end{equation}
Taking the negative logarithm of \eqref{eqPF-2}, removing the constant term, and scaling by $\frac{2}{T}$, we get
\begin{equation}\label{eqPF-3}
f(\boldsymbol{x}) \triangleq \frac{1}{T} \sum_{t=1}^T [\boldsymbol{\tilde{v}}(t)-\boldsymbol{\tilde{v}}(t,\boldsymbol{x})]^T \Sigma^{-1}_n [\boldsymbol{\tilde{v}}(t)-\boldsymbol{\tilde{v}}(t,\boldsymbol{x})]
\end{equation}

It will be shown in Lemma 1 that the correct phase connection $\boldsymbol{x^*}$ maximizes the likelihood function \eqref{eqPF-2} and minimizes $f(x)$ under two mild assumptions.

\begin{lemma} \label{Lemma_joint}
Let $\boldsymbol{x^*}$ be the correct phase connection. If the following two conditions are satisfied, then as $T\rightarrow \infty$, $\boldsymbol{x^*}$ is a global optimizer to minimize $f(\boldsymbol{x})$.
\begin{enumerate}
\item $\boldsymbol{n}(t_k)$ is i.i.d. and independent of $\boldsymbol{\tilde{v}}^{\text{ref}}(t_l)$, $\boldsymbol{\tilde{p}}(t_l)$, and $\boldsymbol{\tilde{q}}(t_l)$, for $\forall t_k,t_l \in Z^+$.
\item $\boldsymbol{\tilde{v}}^{\text{ref}}(t_k)$, $\boldsymbol{\tilde{p}}(t_k)$, and $\boldsymbol{\tilde{q}}(t_k)$ are independent of $\boldsymbol{\tilde{v}}^{\text{ref}}(t_l)$, $\boldsymbol{\tilde{p}}(t_l)$, and $\boldsymbol{\tilde{q}}(t_l)$, for $\forall t_k,t_l\in Z^+, \ t_k \neq t_l$
\end{enumerate}
\end{lemma}

The proof of Lemma 1 can be found in Appendix \ref{App_Lemma_joint}. By substituting \eqref{eqPF-1} into \eqref{eqPF-3}, we can see that directly minimizing $f(\boldsymbol{x})$ is very difficult due to its nonlinearity and nonconvexity. Furthermore, the actual value of $\Sigma_n$ is unknown. To address this technical challenge, in Section \ref{subsection_MMLE}, we will convert the phase identification problem into an MMLE problem and prove that the correct phase connection is also a global optimizer of the MMLE problem.


\subsection{MMLE Problem Formulation} \label{subsection_MMLE}
Let $\tilde{v}_m(t)$ be the $m$th entry of $\boldsymbol{\tilde{v}}(t)$, $\tilde{v}_m(t,\boldsymbol{x})$ be the $m$th entry of $\boldsymbol{\tilde{v}}(t,\boldsymbol{x})$, and $n_m(t)$ be the $m$th entry of $\boldsymbol{n}(t)$. The marginal likelihood of observing $\{\tilde{v}_m(t)\}^{T}_{t=1}$ given $\{\boldsymbol{\tilde{v}}^{\text{ref}}(t)\}^{T}_{t=1}$, $\{\boldsymbol{\tilde{p}}(t)\}^{T}_{t=1}$, and $\{\boldsymbol{\tilde{q}}(t)\}^{T}_{t=1}$ is a function of $\boldsymbol{x}$:
\begin{equation}\label{eqPF-4}
\begin{aligned}
& Prob(\{\tilde{v}_m(t)\}^{T}_{t=1}| \{\boldsymbol{\tilde{v}}^{\text{ref}}(t)\}^{T}_{t=1}, \{\boldsymbol{\tilde{p}}(t)\}^{T}_{t=1}, \{\boldsymbol{\tilde{q}}(t)\}^{T}_{t=1};\boldsymbol{x}) \\
& =\frac{\Sigma_n(m,m)^{-\frac{T}{2}}}{(2 \pi)^{\frac{T}{2}}} \exp \Big\{\!-\frac{1}{2} \sum_{t=1}^T \frac{[\tilde{v}_m(t)\!-\!\tilde{v}_m(t,\boldsymbol{x})]^2}{\Sigma_n(m,m)} \! \Big\}
\end{aligned}
\end{equation}
where $\Sigma_n(m,m)$ is the $m$th diagonal entry of $\Sigma_n$. Taking the negative logarithm of \eqref{eqPF-4}, removing the constant term, and scaling by $\frac{2 \Sigma_n(m,m)}{T}$, we have
\begin{equation}\label{eqPF-5}
f_m(\boldsymbol{x}) \triangleq \frac{1}{T} \sum_{t=1}^T [\tilde{v}_m(t)-\tilde{v}_m(t,\boldsymbol{x})]^2
\end{equation}

\begin{lemma} \label{Lemma_marginal}
Let $\boldsymbol{x^*}$ be the correct phase connection. If the two conditions in Lemma \ref{Lemma_joint} hold, then $\boldsymbol{x^*}$ is a global optimizer to minimize $f_m(\boldsymbol{x})$ as $T\rightarrow \infty$. In addition, any $\boldsymbol{x}$ is a global optimizer of $f_m(\boldsymbol{x})$ if it satisfies all the following conditions:
\begin{enumerate}
\item $x_m^i={x^*}_m^i, \forall i$;
\item $x_k^i={x^*}_k^i, \forall i$, $k \neq m$ and load $k$ is not three-phase.
\end{enumerate}
\end{lemma}

The proof of Lemma 2 can be found in Appendix \ref{App_Lemma_marginal}. 



\subsection{Solution Method for the MMLE Problem} \label{sec_solve_MMLE}
Directly minimizing $f_m(\boldsymbol{x})$ from \eqref{eqPF-5} is still a difficult task. Thus, we further simplify the optimization problem by first solving three subproblems $min f_{m,i}(\boldsymbol{x}_{-m}), i \in \{1, 2, 3\}$. 
$f_{m,i}(\boldsymbol{x}_{-m})$ are defined as
\begin{equation}\label{eqPF-6}
\begin{aligned}
& f_{m,i}(\boldsymbol{x}_{-m}) \triangleq f_m(\boldsymbol{x}) \\
\text{subject to}  \ & x_m^i=1 \ \text{and} \ x_m^j=0 \ \text{for} \ j \neq i
\end{aligned}
\end{equation}
where $\boldsymbol{x}_{-m}$ is a $(3M-3)\times 1$ vector containing every element in $\boldsymbol{x}$ except $x_m^1$, $x_m^2$, and $x_m^3$. Since $x_m^i=0 \ \text{or} \ 1$, and $\sum_i x_m^i=1$, then from \eqref{eqPF-6} we have: 
\begin{equation} \label{eqProp_min}
\begin{aligned}
 \min_{\boldsymbol{x}} f_m(\boldsymbol{x})= & \min_{i=1,2,3} \ \min_{\boldsymbol{x}_{-m}} f_{m,i}(\boldsymbol{x}_{-m})
\end{aligned}
\end{equation}

To solve the sub-problems, we first define $\tilde{v}_{m,i}(t,\boldsymbol{x}_{-m})$ as
\begin{equation}\label{eqPF-7}
\begin{aligned}
& \tilde{v}_{m,i}(t,\boldsymbol{x}_{-m}) \triangleq \tilde{v}_m(t,\boldsymbol{x}) \\
\text{subject to}  \ & x_m^i=1 \ \text{and} \ x_m^j=0 \ \text{for} \ j \neq i
\end{aligned}
\end{equation}
Substituting \eqref{eqPF-1} into \eqref{eqPF-7}, we have
\begin{equation}\label{eqPF-8}
\begin{aligned}
\tilde{v}_{m,i}(t,\boldsymbol{x}_{-m})  = & \tilde{v}^{\text{ref}}_{m,i}(t)+\hat{K}_{m,i}X^T\boldsymbol{\tilde{p}}(t)+\hat{L}_{m,i}X^T\boldsymbol{\tilde{q}}(t)
 \\
\text{subject to} \ & x_m^i=1 \ \text{and} \ x_m^j=0 \ \text{for} \ j \neq i
\end{aligned}
\end{equation}
where $\tilde{v}^{\text{ref}}_{m,i}(t)$ is the entry of $\boldsymbol{\tilde{v}}^{\text{ref}}(t)$ corresponding to $x_m^i$, $\hat{K}_{m,i}$ and $\hat{L}_{m,i}$ are the row vectors of $\hat{K}$ and $\hat{L}$ corresponding to $x_m^i$. 

Define an $M \times 3M$ matrix $\mathfrak{D}$ as:
\begin{equation}\label{eqPF-9}
\mathfrak{D}\triangleq
\textmd{diag}(\underbrace{[1 \ 1 \ 1],\ldots , [1 \ 1 \ 1]}_{\text{repeat } M \text{times}})
\end{equation}

Then matrix $X$ can be expressed by decision vector $\boldsymbol{x}$ as $X=\mathfrak{D} \ \textmd{diag}(\boldsymbol{x})$. Thus, we can simplify the second term on the right-hand-side (RHS) of \eqref{eqPF-8} as
\begin{equation}\label{eqPF-11}
\begin{aligned}
& \hat{K}_{m,i}X^T\boldsymbol{\tilde{p}}(t)=\hat{K}_{m,i} \ \textmd{diag}(\boldsymbol{x}) \ \mathfrak{D}^T \ \boldsymbol{\tilde{p}}(t) \\
= & \boldsymbol{x}^T \ \textmd{diag}(\hat{K}_{m,i}) \ \mathfrak{D}^T \ \boldsymbol{\tilde{p}}(t) = \boldsymbol{x}^T \ \boldsymbol{\zeta}_{m,i}(t) = \boldsymbol{\zeta}_{m,i}^T(t) \ \boldsymbol{x}
\end{aligned}
\end{equation}
where $\boldsymbol{\zeta}_{m,i}(t) \triangleq \textmd{diag}(\hat{K}_{m,i}) \ \mathfrak{D}^T \ \boldsymbol{\tilde{p}}(t)$. Similarly, simplify the third term on the RHS of \eqref{eqPF-8} as
\begin{equation}\label{eqPF-12}
\begin{aligned}
& \hat{L}_{m,i}X^T\boldsymbol{\tilde{q}}(t) = \boldsymbol{\xi}_{m,i}^T(t) \ \boldsymbol{x}
\end{aligned}
\end{equation}
where $\boldsymbol{\xi}_{m,i}(t) \triangleq \textmd{diag}(\hat{L}_{m,i}) \ \mathfrak{D}^T \ \boldsymbol{\tilde{q}}(t)$.

Substituting \eqref{eqPF-11} and \eqref{eqPF-12} into equation \eqref{eqPF-8}, we have

\begin{equation}\label{eqPF-13}
\begin{aligned}
& \tilde{v}_m(t)-\tilde{v}_{m,i}(t,\boldsymbol{x}_{-m}) \\
= & \tilde{v}_m(t)-\tilde{v}^{\text{ref}}_{m,i}(t)-\boldsymbol{\zeta}_{m,i}^T(t) \boldsymbol{x}-\boldsymbol{\xi}_{m,i}^T(t) \boldsymbol{x} \\
= & \tilde{v}_m(t)-\tilde{v}^{\text{ref}}_{m,i}(t)-\boldsymbol{\psi}_{m,i}^T(t) \boldsymbol{x}\\
= & \tilde{v}_m(t)-\tilde{v}^{\text{ref}}_{m,i}(t)-[\boldsymbol{\varphi}_{m,i}^T(t) \boldsymbol{x}_{-m}+\eta_{m,i}(t)]\\
= & v^{\text{tot}}_{m,i}(t)-\boldsymbol{\varphi}^T_{m,i}(t) \boldsymbol{x}_{-m}
\end{aligned}
\end{equation}
Where $\boldsymbol{\psi}_{m,i}(t) \triangleq \boldsymbol{\zeta}_{m,i}(t) + \boldsymbol{\xi}_{m,i}(t)$. $\boldsymbol{\varphi}_{m,i}(t)$ is a vector containing all the elements in $\boldsymbol{\psi}_{m,i}(t)$ except the three elements corresponding to $x_m^1$, $x_m^2$, and $x_m^3$. $\eta_{m,i}(t)$ is the element in $\boldsymbol{\psi}_{m,i}(t)$ corresponding to $x_m^i$. In the last line of \eqref{eqPF-13}, $v^{\text{tot}}_{m,i}(t)$ is defined as $v^{\text{tot}}_{m,i}(t) \triangleq \tilde{v}_m(t)-\tilde{v}^{\text{ref}}_{m,i}(t)-\eta_{m,i}(t)$. 

Note that our proposed phase identification method still works even if there is a topology change in the primary feeder. If such topology change occurs at time $t_c$, then we can simply update $v^{\text{tot}}_{m,i}(t)$ and $\boldsymbol{\varphi}_{m,i}(t)$ in \eqref{eqPF-13} according to the new primary feeder topology.

With \eqref{eqPF-13}, the function $f_{m,i}(\boldsymbol{x}_{-m})$ can be transformed into
\begin{equation}\label{eqPF-14}
\begin{aligned}
f_{m,i}(\boldsymbol{x}_{-m})=\frac{1}{T} \sum_{t=1}^T [v^{\text{tot}}_{m,i}(t)-\boldsymbol{\varphi}^T_{m,i}(t) \boldsymbol{x}_{-m}]^2
\end{aligned}
\end{equation}
Now each MMLE sub-problem in \eqref{eqProp_min} can be formulated as
\begin{equation}\label{eqPF-15}
\begin{aligned}
& \text{find} & \boldsymbol{x}^{\dag}_{-m,i}=\argmin_{\boldsymbol{x}_{-m}} \ f_{m,i}(\boldsymbol{x}_{-m}) \\
&\text{subject to} & x_k^j=0 \ \text{or} \ 1 \quad \forall j \ \text{and} \ k \neq m \\
&                 & \sum_j x_k^j=1 \quad \forall k \neq m.
\end{aligned}
\end{equation}
This is a binary least-square problem. To solve it efficiently, we can further relax the problem by replacing the binary constraint by its convex hull. Now the problem is equivalent to convex quadratic programming, which can be solved in polynomial time \cite{Vavasis2001}. The continuous solution of $\boldsymbol{x}_{-m}$ in the convex hull can then be rounded to binary values as follows: for each load $k\neq m$, round $x_{k}^j$ to $1$ if it is the largest among $x_{k}^1$, $x_{k}^2$, and $x_{k}^3$, and round the other two variables to 0.




\subsection{Phase Identification Algorithm}
Our proposed MMLE-based phase identification algorithm is summarized in Algorithm \ref{Algorithm_phase} and explained as follows. From step 1 to 6, we solve $M$ MMLE problems, each of which contains three binary least-square sub-problems. Step 3 solves the sub-problems of MMLE based on \eqref{eqPF-15}. Based on \eqref{eqProp_min}, step 5 solves the $m$th MMLE problem by finding which of the three $\boldsymbol{x}^{\dag}_{-m,i} \ (i=1,2,3)$ minimizes $f_m(\boldsymbol{x})$. The chosen $\boldsymbol{x}^{\dag}_{-m,i}$, combined with the corresponding $x_{m}^i=1$ and $x_{m}^j=0$ ($j\neq i$), forms the $3M \times 1$ solution $\boldsymbol{x^{\dag}}_m$ of the $mth$ MMLE problem. The $M$ sets of $\boldsymbol{x^{\dag}}_m$ may not be all correct due to the limited number of measurements and measurement noise. Thus, in step 7, we design two approaches to integrate $M$ sets of $\boldsymbol{x^{\dag}}_m$ into two phase identification solutions:



\begin{enumerate}
\item \textit{Target-only Approach}. The phase connection of each load $m$ is the corresponding connection shown in the $m$th solution $\boldsymbol{x^{\dag}}_m$.
\item \textit{Voting Approach}. For a single-phase or two-phase load $m$, the phase connection is the corresponding phase connection that receives the most votes in the $M$ sets of $\boldsymbol{x^{\dag}}_m$. For a three-phase load $m$, the phase connection is still determined by the target-only approach. 
\end{enumerate} 
In step 8, we calculate $\sum_{m=1}^M f_m(\boldsymbol{x})$ based on the phase identification solution of both the target-only and the voting approaches. The final phase identification solution is the one that has the lower sum of square error.

\begin{algorithm}[htb]
\caption{Phase Identification Algorithm} \label{Algorithm_phase}
\begin{algorithmic}[1]
\Require $\boldsymbol{\tilde{v}}(t)$, $\boldsymbol{\tilde{v}}^{\text{ref}}(t)$, $\boldsymbol{\tilde{p}}(t)$, $\boldsymbol{\tilde{q}}(t)$, $\hat{K}$, and $\hat{L}$, $t=1,...,T$.
\Ensure Estimated phase connections for the $M$ loads.
        \For{$m=1$ to $M$}
           \For{$i=1$ to $3$}
               \State Use the input to calculate $v^{\text{tot}}_{m,i}(t)$ and $\boldsymbol{\varphi}^T_{m,i}(t)$ and find the solution $\boldsymbol{x}^{\dag}_{-m,i}$ to the sub-problem in \eqref{eqPF-15}. 
           \EndFor
           \State Use $\boldsymbol{x}^{\dag}_{-m,i}$, $i\in\{1,2,3\}$ to find the $\boldsymbol{x}$ that minimizes $f_m(\boldsymbol{x})$ in \eqref{eqPF-5}. Record the solution as $\boldsymbol{x^{\dag}}_m$. 
        \EndFor
        \State Generate two phase identification results based on $M$ sets of $\boldsymbol{x^{\dag}}_m$ using two approaches: the target-only approach and the voting approach.
        \State Calculate $\sum_{m=1}^M f_m(\boldsymbol{x})$ based on both the target-only and the voting approach. Select the solution with the lower sum of square error.
\end{algorithmic}
\end{algorithm}

\section{Numerical Study} \label{Sec_numerical}
\subsection{Setup for Numerical Tests}
The performance of our proposed MMLE-based algorithm is evaluated using the IEEE 37-bus, 123-bus, and 342-bus test circuits. The results will show that the proposed algorithm works well for distribution networks with either tree structured feeders (37-bus and 123-bus) or heavily meshed primary feeders (342-bus). To make the task more difficult, we modify the test feeders to include all possible phase connection types (single-phase, two-phase, and three-phase). The number of loads by phase connection type is summarized in Table \ref{table_phase_alloc}. Fig. \ref{fig_123_node} illustrates the schematic of the 123-bus circuit.


\begin{table}[htb]
\centering
\caption{Number of Loads Per Phase in the IEEE Test Circuits} 
\label{table_phase_alloc}
\begin{tabular}{ c  c  c  c  c  c  c  c  c }
  \hline
  \hline
  Feeder & A & B & C & AB & BC & CA & ABC & Total\\ \hline
  37-bus & 5 & 5 & 6 & 3 & 2 & 2 & 2 & 25 \\ 
  123-bus & 18 & 17 & 17 & 9 & 9 & 10 & 5 & 85 \\
  342-bus & 30 & 38 & 31 & 35 & 31 & 33 & 10 & 208 \\ \hline \hline
\end{tabular}
\end{table}

The hourly average real power consumption measurements from smart meters of a distribution feeder managed by FortisBC are used in test feeders. The length of the real power consumption time series is 2160, which represents 90 days of hourly smart meter measurements. The reactive power time series are generated by randomly sampling power factors from a uniform distribution $\mathcal{U}(0.9,1)$ to represent lagging loads. The peak loads for the three IEEE test circuits are 2.4 MW, 4 MW, and 43 MW. The power flows of the test circuits are simulated using OpenDSS. All smart meter measurements contain noise that follows zero-mean Gaussian distributions with three-sigma deviation matching 0.1\% to 0.2\% of the nominal values. The 0.1 and 0.2 accuracy class smart meters established in ANSI C12.20-2015 are typical in real-world implementations. To make the phase identification task even more challenging, we assume that older generations of smart meters are adopted. That is to say, after adding measurement noise, the voltage measurements are rounded to the nearest 1 V for primary line loads and 0.1 V for secondary loads. The real and reactive power measurements are rounded to the nearest 0.1 kW or 0.1 kVAr. The relaxed optimization problems in equation \eqref{eqPF-15} are solved using CPLEX on a DELL workstation with 3.3 GHz Intel Xeon CPU and 16 GB of RAM.


Before presenting the main numerical results, we first verify the Gaussianity assumption for the noise term $\boldsymbol{n}(t)$ in equation \eqref{eqML-6}. The Kolmogorov-Smirnov test is used to verify the Gaussianity assumption. With a significance level of 5\%, the noise terms for all loads pass the test except 9 loads at 0.1\% meter accuracy level and 1 load at 0.2\% meter accuracy level in the 342-bus circuit. By checking the normalized auto-correlations of $n(t)$, we found the noise to be uncorrelated over time. For Gaussian random variables, this indicates independence over time.

\subsection{Performance of the Proposed Phase Identification Method}
The phase identification accuracy of our proposed MMLE-based algorithm is shown in Table \ref{table_accuracy}, which covers three IEEE test feeders, two meter accuracy classes (0.1\% and 0.2\%), and three time windows (30 days, 60 days, 90 days). With 90 days of hourly meter measurements and both accuracy class meters, the proposed algorithm achieved 100\% accuracy for all three IEEE distribution test circuits. The proposed algorithm works well not only for radial feeders (37-bus, 123-bus), but also the meshed circuit (342-bus). As shown in the table, the accuracy of the MMLE-based phase identification algorithm increases as the smart meter measurement error decreases. When additional smart meter data becomes available, the phase identification accuracy of the proposed algorithm also increases as expected. The average computation time of the algorithm with 90 days of data is only around 1.3 seconds, 6.5 seconds, and 256 seconds for the three circuits, respectively.

\begin{table}[htb]
\centering
\caption{Accuracy of the Proposed Phase Identification Method} 
\label{table_accuracy}
\begin{tabular}{  c c  c  c  c}
  \hline
  \hline
  Feeder & Meter Class & 30 Days & 60 Days & 90 Days\\ \hline
  37-bus & 0.1\% & 100\% & 100\% & 100\% \\ 
  & 0.2\% & 92\% & 100\% & 100\% \\
  \hline
  123-bus & 0.1\% & 96.47\% & 100\% & 100\% \\ 
  & 0.2\% & 63.53\% & 96.47\% & 100\% \\
  \hline
  342-bus & 0.1\% & 96.63\% & 100\% & 100\% \\ 
  & 0.2\% & 72.60\% & 99.52\% & 100\% \\ 
  \hline \hline
\end{tabular}
\end{table}
\subsection{Comparison With Existing Methods}
The phase identification accuracy of our proposed MMLE-based method is compared with two state-of-the-art methods: the correlation-based approach \cite{xu2018phase} and the clustering-based approach \cite{wang2017advanced}. We also evaluate the robustness of the phase identification algorithms with respect to inaccurate feeder models and incomplete measurements.

The 123-bus and 342-bus test feeders with 90 days of 0.1\% accuracy class smart meter measurements are used for the comparison. To introduce incomplete smart meter measurements, we gradually decrease the penetration ratio of smart meters from 100\% to 10\% with a 10\% step. To create inaccurate feeder models, we introduce noisy network parameters and inaccurate topology information. Specifically, we add zero-mean Gaussian noise with three-sigma deviation matching 30\% of the nominal values to the actual line admittance of the 123-bus and 342-bus feeders. Eight secondary branches are assumed to be missing in the topology model of the 342-bus feeder.

Note that the correlation-based method \cite{xu2018phase} was originally designed to handle single-phase loads only. Thus, we extend it to accommodate two-phase loads. To make it a fair comparison, we assume that the information of whether a particular load is one-phase, two-phase, or three-phase is known to all algorithms. Inaccurate feeder models and incomplete measurements do not affect the correlation-based and clustering-based algorithms directly. This is because these two methods do not rely on the primary feeder model. Similarly, the MMLE-based method simply constructs a formulation with a smaller decision vector $\boldsymbol{x}$ when dealing with incomplete meter measurements.

\begin{figure}[htb]
\centering
    \begin{subfigure}{0.24\textwidth}
        \includegraphics*[width=\textwidth]{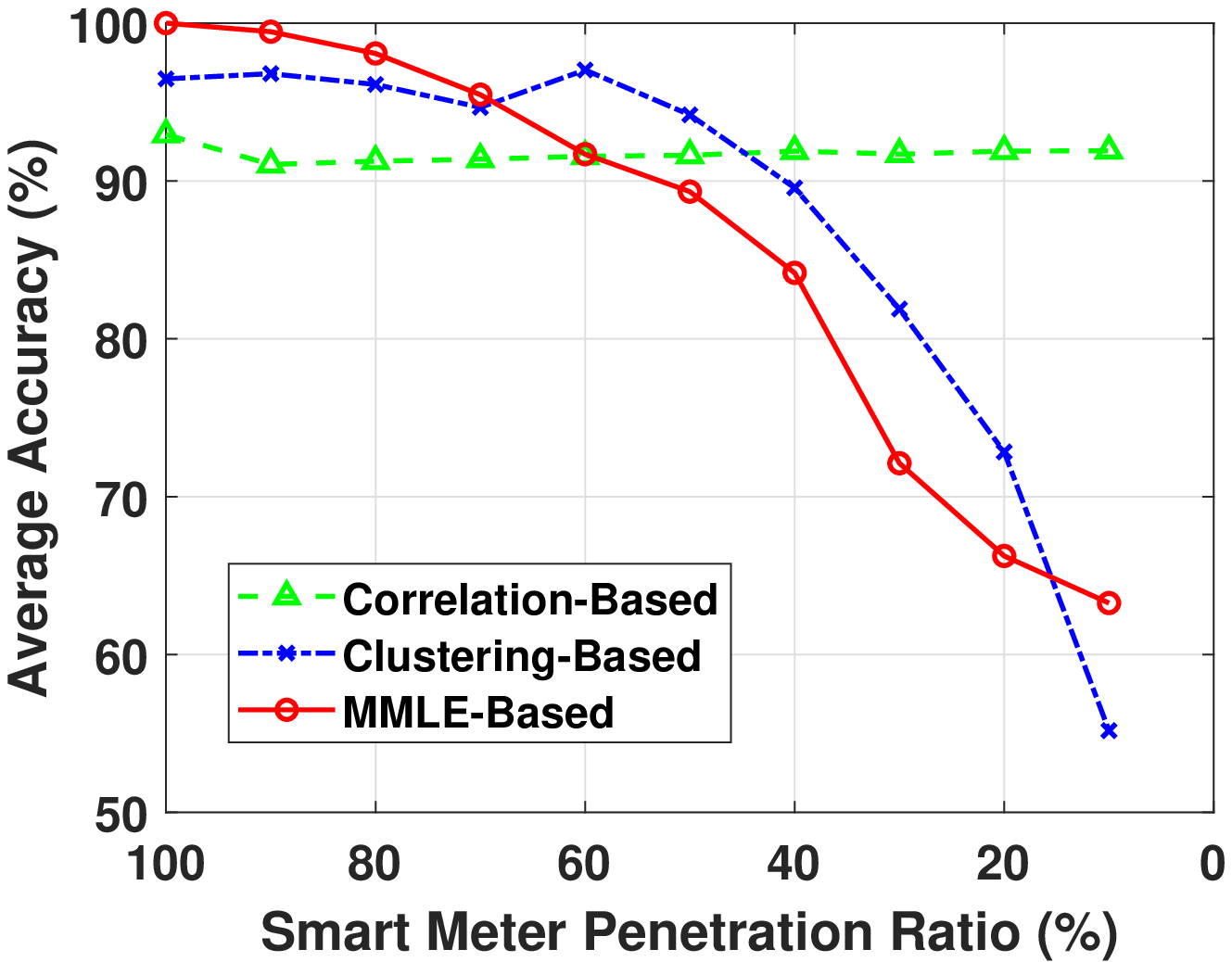}
        \caption{123-bus}
        \label{fig_sub_123}
    \end{subfigure}
    \begin{subfigure}{0.24\textwidth}
        \includegraphics*[width=\textwidth]{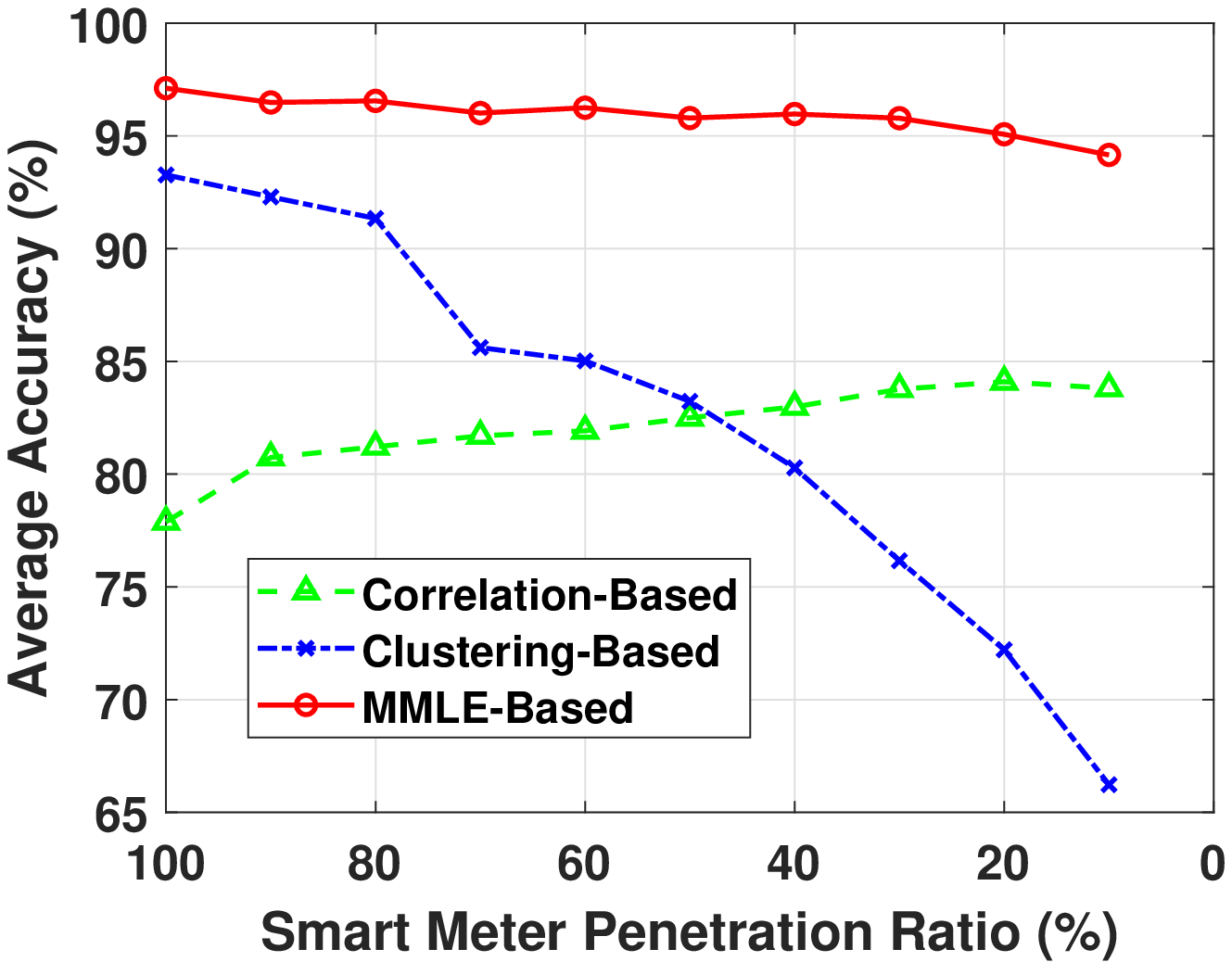}
        \caption{342-bus}
        \label{fig_sub_342}
    \end{subfigure}
\caption{Phase identification accuracy of three methods}
\label{fig_compare}
\end{figure}

The average phase identification accuracies of the proposed algorithm and two benchmark algorithms with different smart meter penetration ratios and inaccurate feeder models are shown in Fig. \ref{fig_compare}. When the smart meter penetration rate is not 100\%, we randomly select the location of smart meters around 50 times and calculate the average accuracies. 

As shown in Fig.\ref{fig_compare}, our proposed MMLE-based algorithm achieves around 97\% accuracy on the 342-bus feeder at the 100\% smart meter penetration rate. This is lower than the 100\% accuracy reported in Table \ref{table_accuracy} due to an inaccurate primary feeder model. Our proposed algorithm yields higher accuracy for the 123-bus radial feeder when the smart meter penetration rate is at 70\% or higher. For the more complex 342-bus feeder, which is heavily meshed, our proposed algorithm outperforms both existing algorithms across all smart meter penetration levels. Our proposed algorithm is more robust with respect to incomplete measurements on the heavily meshed 342-bus feeder than on the radial 123-bus feeder. To explain this phenomenon, we examine the sensitivity of $\tilde{v}_m(t,\boldsymbol{x})$, the smart meter voltage measurement for load $m$, with respect to the phase connection decision vector $\boldsymbol{x}$. It turns out that in the 342-bus feeder, load $m$'s voltage measurement is more sensitive to its own phase connection decision variables and less sensitive to the phase connection decision variables of other loads. As the penetration level of smart meters continue to increase around the world, the comparative advantage of our proposed algorithm will become more pronounced.

\section{Conclusion} \label{Sec_conclusion}

This paper develops a physically inspired data-driven algorithm for the phase identification in power distribution systems. The phase identification problem is first formulated as an MLE and MMLE problem based on the three-phase power flow manifold. We prove that the correct phase connection is a global optimum for both the MLE and the MMLE problems. A computationally efficient algorithm is developed to solve the MMLE problem, which involves synthesizing the solutions from the sub-problems via the voting and the target-only approaches. The sub-problems are further transformed into an equivalent binary least square form and solved efficiently by relaxing the binary constraints. Comprehensive simulation results with real-world smart meter data and IEEE distribution test circuits show that our proposed phase identification algorithm yields high accuracy and outperforms existing methods. The proposed algorithm is also fairly robust with respect to inaccurate feeder models and incomplete measurements. 





\bibliographystyle{IEEEtran}
\bibliography{reference_PF_phase_ID}

\clearpage 
\appendices

\section{Simplification of Single-Phase and Two-Phase Branches} \label{1p2pEst}

To convert loaded single-phase and two-phase branches into a load directly connected to the primary feeder, we need to estimate each branch's equivalent power injection and voltage magnitude. In other words, given the line impedances of single-phase and two-phase branches, the voltage magnitudes and power injections of the loads, we need to calculate the equivalent power injection and voltage magnitude on the primary feeder. The conversion of single-phase and two-phase branches is carried out separately below.

\subsubsection{Simplification of a Single-Phase Line}
Suppose there is a single-phase line with impedance $z$ serving a load with power injection $S$ and voltage magnitude $|V|$. It is assumed that the power injection $S$ and the voltage magnitude $|V|$ are given. Thus, the current injection magnitude $|I|$ and power factor angle $\phi$ can be calculated. Then, at the upstream port of the primary feeder, the single-phase line's equivalent voltage magnitude is $||V|-z|I|\angle\!-\!\phi|$ and the equivalent power injection is $S-z|I|^2$.


\subsubsection{Simplification of a Two-Phase Line}
For a two-phase line serving a load, the voltage drop along the  line section can be described by
\begin{equation}\label{eqLE-2}
\begin{bmatrix}
V_n^1 \\
V_n^2
\end{bmatrix}
=
\begin{bmatrix}
z_{11} & z_{12} \\
z_{21} & z_{22}
\end{bmatrix}
\begin{bmatrix}
-I \\
I
\end{bmatrix}
+
\begin{bmatrix}
V_m^1 \\
V_m^2
\end{bmatrix}
\end{equation}
where $z_{11}$, $z_{12}$, $z_{21}$, and $z_{22}$ form the line impedance matrix, which is assumed to be known. $V_n^1$, $V_n^2$, $V_m^1$, and $V_m^2$ are the nodal voltage phasors of the upstream port and the load, which are assumed to be unknown. $I$ is the current injection phasor of the load. Subtracting row 2 from row 1 in \eqref{eqLE-2}, we have
\begin{equation}\label{eqLE-3}
V^{12}_n=(z_{12}+z_{21}-z_{11}-z_{22})I+V^{12}_m=z_{sum}I+V^{12}_m
\end{equation}
where $V^{12}_n=V_n^1-V_n^2$ and $V^{12}_m=V_m^1-V_m^2$. For load $m$, using the measured voltage magnitude $|V^{12}_m|$ and power injection $S_m$, we calculate the current injection magnitude $|I|$ and the power factor angle $\phi$. Then, at the upstream port of the primary feeder, the two-phase line's equivalent voltage magnitude is $||V^{12}_m|+z_{sum}|I|\angle\!-\!\phi|$ and the equivalent power injection is $S_m+z_{sum}|I|^2$.



\section{Derivation of the Transformed Linearized Three-phase Power Flow Model} \label{NS_derive} 
Let $A_{mn}^{ij}$ be the $(N+1)\times(N+1)$ block in matrix $A_{mn}$ corresponding to phase $ij$. Suppose the first row and column of $A_{mn}$ correspond to the substation node, then $A_{mn}^{ij}$ can be divided into 4 blocks as follows:


\begin{equation}\label{eqNS-1}
A_{mn}^{ij}=
\begin{bmatrix}
d_{mn}^{ij} & (\boldsymbol{b}_{mn}^{ij})^T \\
\boldsymbol{b}_{mn}^{ij} & \check{A}_{mn}^{ij}
\end{bmatrix}
\end{equation}
where $\check{A}_{mn}^{ij}$ is a nonsingular $N\times N$ matrix. Define $\check{A}_{mn}$ as the collection of $\check{A}_{mn}^{ij}$ over all $i$ and $j$, $B_{mn}$ as the collection of $\boldsymbol{b}_{mn}^{ij}$ over all $i$ and $j$, $C_{mn}$ as the collection of $(\boldsymbol{b}_{mn}^{ij})^T$ over all $i$ and $j$, and $D_{mn}$ as the collection of $d_{mn}^{ij}$ over all $i$ and $j$. By permuting the variables and corresponding matrix rows and columns, \eqref{eqLM-1} can be transformed into
\begin{equation}\label{eqNS-10}
\begin{bmatrix}
\check{A}_{11} & \check{A}_{12} & B_{11} & B_{12}\\
\check{A}_{21} & \check{A}_{22} & B_{21} & B_{22}\\
C_{11} & C_{12} & D_{11} & D_{12} \\
C_{21} & C_{22} & D_{21} & D_{22} 
\end{bmatrix}
\begin{bmatrix}
\boldsymbol{v}_{-0}-\boldsymbol{\overline{v}}_{-0} \\
\boldsymbol{\theta}_{-0}-\boldsymbol{\overline{\theta}}_{-0} \\
\boldsymbol{v}_0-\boldsymbol{\overline{v}}_0 \\
\boldsymbol{\theta}_0-\boldsymbol{\overline{\theta}}_0
\end{bmatrix}
=
\begin{bmatrix}
\boldsymbol{p}_{-0}\\
\boldsymbol{q}_{-0} \\
\boldsymbol{p}_0 \\
\boldsymbol{q}_0
\end{bmatrix}
\end{equation}
where $(\cdot)_{-0}$ denotes a vector excluding the substation node, and $(\cdot)_0$ denotes a vector of the substation node.

Define Matrix $\mathcal{D}$ as follows:
\begin{equation}\label{eqNS-12}
\mathcal{D}=\textmd{diag}(\mathbf{1}_N,\mathbf{1}_N,\mathbf{1}_N,\mathbf{1}_N,\mathbf{1}_N,\mathbf{1}_N)
\end{equation}
From the property of admittance matrix $Y^{ij}$, we have $A^{ij}_{mn} \mathbf{1}_{N+1}=\mathbb{0}_{(N+1)\times 1}$ and $[\check{A}^{ij}_{mn},\boldsymbol{b}^{ij}_{mn}] \mathbf{1}_{N+1}=\mathbb{0}_{N\times 1}$. 

Thus, we have the following equality relationship:
\begin{equation}\label{eqNS-13}
\begin{bmatrix}
\check{A}_{11} & \check{A}_{12} & B_{11} & B_{12}\\
\check{A}_{21} & \check{A}_{22} & B_{21} & B_{22}
\end{bmatrix}
\begin{bmatrix}
\mathcal{D} \\
I_{6\times 6}
\end{bmatrix}
=\mathbb{0}_{6N\times 6}
\end{equation}
Now, it can be easily shown that
\begin{equation}\label{eqNS-14}
\begin{bmatrix}
B_{11} & B_{12}\\
B_{21} & B_{22}
\end{bmatrix}
=-
\begin{bmatrix}
\check{A}_{11} & \check{A}_{12}\\
\check{A}_{21} & \check{A}_{22}
\end{bmatrix}
\mathcal{D}
\end{equation}
Plugging equation \eqref{eqNS-14} into equation \eqref{eqNS-10}, we have
\begin{equation}\label{eqNS-5}
\begin{bmatrix}
\check{A}_{11} & \check{A}_{12}\\
\check{A}_{21} & \check{A}_{22}
\end{bmatrix}
\begin{bmatrix}
\boldsymbol{v}_{-0}^a - \mathbf{1}_N v_0^a\\
\boldsymbol{v}_{-0}^b - \mathbf{1}_N v_0^b\\
\boldsymbol{v}_{-0}^c - \mathbf{1}_N v_0^c\\
\boldsymbol{\theta}_{-0}^a - \mathbf{1}_N \theta_0^a\\
\boldsymbol{\theta}_{-0}^b - \mathbf{1}_N \theta_0^b\\
\boldsymbol{\theta}_{-0}^c - \mathbf{1}_N \theta_0^c
\end{bmatrix}
=
\begin{bmatrix}
\boldsymbol{p}_{-0}\\
\boldsymbol{q}_{-0}
\end{bmatrix}
\end{equation}
where $\boldsymbol{v}_{-0}^i$ and $\boldsymbol{\theta}_{-0}^i$ denote the phase $i$ variables in $\boldsymbol{v}_{-0}$ and $\boldsymbol{\theta}_{-0}$. $v_0^i$ and $\theta_0^i$ denote the substation's voltage magnitude and angle of phase $i$. \eqref{eqNS-5} is exactly the same as \eqref{eqNS-16}.

\section{Estimation of Nodal Power Injection of a Two-phase Load} \label{App_Delta_power}
Define $I_{ab}$ as the current phasor flowing out of the load's phase $A$ port and into the load's phase $B$ port. Let $I_a$ be the injected current phasor from phase $A$ port, and let $I_b$ be the injected current phasor from phase $B$ port. By definition, we know that $I_a=-I_b=I_{ab}$. Let the angle of $V_{ab}$ be the reference angle, i.e., $V_{ab}=|V_{ab}|\angle 0\degree$, then
\begin{equation}\label{eq2P-2}
\begin{aligned}
S_{ab} &=P_{ab}+jQ_{ab} \\
&=V_{ab}I_{ab}^* \\
&=|V_{ab}|[Re(I_{ab})-jIm(I_{ab})]
\end{aligned}
\end{equation}
Thus,
\begin{equation}\label{eq2P-3}
\begin{aligned}
& Re(I_{ab})=\frac{P_{ab}}{|V_{ab}|} \\
& Im(I_{ab})=-\frac{Q_{ab}}{|V_{ab}|}
\end{aligned}
\end{equation}
When the three-phase voltages are close to balance, the nodal phase-to-neutral power injection can be estimated by the two-phase power injection as follows:
\begin{equation}\label{eq2P-4}
\begin{aligned}
S_a & = V_a I_a^* \\
 & \approx \frac{\sqrt{3}}{3} |V_{ab}| \angle-30\degree \cdot I_{ab}^* \\
 & = \frac{\sqrt{3}}{3} |V_{ab}| \angle-30\degree \bigg(\frac{P_{ab}}{|V_{ab}|}+j \frac{Q_{ab}}{|V_{ab}|}\bigg)\\
 & = \frac{\sqrt{3}}{3} [\cos(-30\degree)+j\sin(-30\degree)] (P_{ab}+jQ_{ab})\\
 & = \bigg( \frac{1}{2}P_{ab} + \frac{\sqrt{3}}{6} Q_{ab} \bigg)+j\bigg( \frac{1}{2} Q_{ab} -  \frac{\sqrt{3}}{6} P_{ab} \bigg)
\end{aligned}
\end{equation}
This is exactly the same as \eqref{eq2P-1-1}. Equation \eqref{eq2P-1-2} can be derived in a similar way.

\section{Link the Voltage Magnitude Measurements of Two-Phase Loads to Nodal Values in the Power Flow Model} \label{App_2phase}
In the following derivations, the voltages are in per unit and angles are in radian. For a two-phase load $m$ across phase $ij$ ($ij \in \{ab,bc,ca\}$) at node $n$, we have
\begin{equation}\label{eq2P-6}
\hat{v}_m=v^{ij}_n=\sqrt{(v^i_n)^2+(v^j_n)^2-2v^i_n v^j_n \cos \theta^{ij}_n}
\end{equation}
where $\hat{v}_m$ is load $m$'s magnitude measurement, $v^{ij}_n$ is the voltage magnitude between phase $ij$ at node $n$, $v^i_n$ is the voltage of phase $i$ at node $n$, and $\theta^{ij}_n$ is the voltage phase angle between phase $ij$ at node $n$. 

Similarly, at the substation, we also have
\begin{equation}\label{eq2P-6-2}
v^{ij}_0=\sqrt{(v^i_0)^2+(v^j_0)^2-2v^i_0 v^j_0 \cos \theta^{ij}_0}
\end{equation}
where $v^{ij}_0$, $v^i$, and $\theta^{ij}_0$ are the corresponding nodal values at the substation. Under normal operating conditions, $v_n^i \approx v_n^j \approx 1$, $\theta_n^{ij} \approx \frac{2 \pi}{3}$. From \eqref{eq2P-6} we have
\begin{equation}\label{eq2P-7}
\frac{\partial v^{ij}_n}{\partial v_n^i} \approx \frac{\sqrt{3}}{2}, \quad \frac{\partial v^{ij}_n}{\partial v_n^j} \approx \frac{\sqrt{3}}{2}, \quad \frac{\partial v^{ij}_n}{\partial \theta_n^{ij}} = \frac{\partial v^{ij}_n}{\partial (\theta_n^i-\theta_n^j)} \approx \frac{1}{2}
\end{equation}
Under normal operating conditions, voltage and angle differences between non-substation nodes and the substation node is very small. Thus, we can easily derive \eqref{eq2P-10} from \eqref{eq2P-7} to approximate $\hat{v}_m-v^{ij}_0$.

\section{Proof of Lemma 1} \label{App_Lemma_joint} 
\begin{IEEEproof}
By definition, $\boldsymbol{\tilde{v}}(t)=\boldsymbol{\tilde{v}}(t,\boldsymbol{x^*})+\boldsymbol{n}(t)$. Plugging it into equation \eqref{eqPF-3}, we have
\begin{equation}\label{Proof_joint-1}
\begin{aligned}
& \lim_{T \to \infty} f(\boldsymbol{x})\\
= & \lim_{T \to \infty} \frac{1}{T} \sum_{t=1}^T [\boldsymbol{\tilde{v}}(t,\boldsymbol{x^*})-\boldsymbol{\tilde{v}}(t,\boldsymbol{x})+\boldsymbol{n}(t)]^T \Sigma^{-1}_n \\
& [\boldsymbol{\tilde{v}}(t,\boldsymbol{x^*})-\boldsymbol{\tilde{v}}(t,\boldsymbol{x})+\boldsymbol{n}(t)] \\
= & \lim_{T \to \infty} \frac{1}{T} \sum_{t=1}^T [\boldsymbol{\tilde{v}}(t,\boldsymbol{x^*})-\boldsymbol{\tilde{v}}(t,\boldsymbol{x})]^T \Sigma^{-1}_n [\boldsymbol{\tilde{v}}(t,\boldsymbol{x^*})-\boldsymbol{\tilde{v}}(t,\boldsymbol{x})] \\
& + \lim_{T \to \infty} \frac{2}{T} \sum_{t=1}^T [\boldsymbol{\tilde{v}}(t,\boldsymbol{x^*})-\boldsymbol{\tilde{v}}(t,\boldsymbol{x})]^T \Sigma^{-1}_n \boldsymbol{n}(t) \\
& + \lim_{T \to \infty} \frac{1}{T} \sum_{t=1}^T \boldsymbol{n}(t)^T \Sigma^{-1}_n \boldsymbol{n}(t) \\
\geq & \lim_{T \to \infty} \frac{1}{T} \sum_{t=1}^T \boldsymbol{n}(t)^T \Sigma^{-1}_n \boldsymbol{n}(t)
\end{aligned}
\end{equation}
It should be noted that $\lim_{T \to \infty} \frac{1}{T} \sum_{t=1}^T [\boldsymbol{\tilde{v}}(t,\boldsymbol{x^*})-\boldsymbol{\tilde{v}}(t,\boldsymbol{x})]^T \Sigma^{-1}_n [\boldsymbol{\tilde{v}}(t,\boldsymbol{x^*})-\boldsymbol{\tilde{v}}(t,\boldsymbol{x})] \geq 0$ because $\Sigma^{-1}_n \succ 0$. As stated in condition 1 of Lemma \ref{Lemma_joint}, $\boldsymbol{n}(t)$ is independent of $\boldsymbol{\tilde{v}}(t,\boldsymbol{x})$ and $\boldsymbol{\tilde{v}}(t,\boldsymbol{x^*})$, so we have $E([\boldsymbol{\tilde{v}}(t,\boldsymbol{x^*})-\boldsymbol{\tilde{v}}(t,\boldsymbol{x})]^T \Sigma^{-1}_n \boldsymbol{n}(t))=0$. Condition 1 and 2 of Lemma \ref{Lemma_joint} also make $[\boldsymbol{\tilde{v}}(t,\boldsymbol{x^*})-\boldsymbol{\tilde{v}}(t,\boldsymbol{x})]^T \Sigma^{-1}_n \boldsymbol{n}(t)$ a sequence of independent variables. Under normal system operating conditions, $[\boldsymbol{\tilde{v}}(t,\boldsymbol{x^*})-\boldsymbol{\tilde{v}}(t,\boldsymbol{x})]^T \Sigma^{-1}_n \boldsymbol{n}(t)$ has limited variance. By Kolmogorov's Strong Law of Large Numbers \cite{Greene11}, $\lim_{T \to \infty} \frac{2}{T} \sum_{t=1}^T [\boldsymbol{\tilde{v}}(t,\boldsymbol{x^*})-\boldsymbol{\tilde{v}}(t,\boldsymbol{x})]^T \Sigma^{-1}_n \boldsymbol{n}(t) \to 0$. Therefore, inequality \eqref{Proof_joint-1} holds. In addition, the minimum of $\lim_{T \to \infty} f(\boldsymbol{x})$ is achieved when $\boldsymbol{x}=\boldsymbol{x^*}$.
\end{IEEEproof}

\section{Proof of Lemma 2} \label{App_Lemma_marginal} 
\begin{IEEEproof}
Following a procedure similar to Appendix \ref{App_Lemma_joint}, we can prove that $\lim_{T \to \infty} f_m(\boldsymbol{x}) \geq \lim_{T \to \infty} \frac{1}{T} \sum_{t=1}^T n_m(t)^2$, and the minimum of $\lim_{T \to \infty} f_m(\boldsymbol{x})$ is achieved when $\boldsymbol{x}=\boldsymbol{x^*}$. Condition 1) and 2) in Lemma \ref{Lemma_marginal} simply mean that we can assign any three-phase loads except load $m$ to any phase and get the same optimum value. This is true, because changing three-phase loads' decision variables does not change the power injections in the system. As long as condition 1) and 2) of Lemma \ref{Lemma_marginal} hold, $\tilde{v}_m(t,\boldsymbol{x})=\tilde{v}_m(t,\boldsymbol{x^*})$. This can also be verified by the structure of $\hat{U}^1$ and $\hat{U}^2$ for three-phase loads. 

\end{IEEEproof}
\end{document}